# Survey of Major Load Balancing Algorithms in Distributed System

Igor N. Ivanisenko
Postgraduate student for Applied Mathematics department
Kharkiv National University of Radioelectronics
14, Lenin ave, Kharkiv, 61166, Ukraine
+38067-777-57-72, ivanisenko79@yahoo.com

Tamara A. Radivilova
Associated professor for Telecommunication Systems department
Kharkiv National University of Radioelectronics
14, Lenin ave, Kharkiv, 61166, Ukraine
tamara.radivilova@gmail.com

*Abstract*— The classification of the most used load balancing algorithms in distributed systems (including cloud technology, cluster systems, grid systems) is described. Comparative analysis of types of the load balancing algorithms is conducted in accordance with the classification, the advantages and drawbags of each type of the algorithms are shown. Performance indicators characterizing each algorithm are indicated.

*Keywords*— load balancing algorithm, distributed system, throughput, resource utilization, capacity

## I. Introduction

Due to a massive spread of distributed computing systems the problem of their effective use has become urgent. One aspect of this problem is the effective planning and task allocation within a distributed computing system in order to optimize resource utilization and reduce the computation time. Quite frequently the situation arises, where a part of the processing resources is idle, while the other part of the resources is overloaded and a large number of tasks are waiting for execution in the queue.

To optimize the use of resources, reduce the time of service requests, horizontal scaling (dynamic addition / removal of devices), as well as to provide fault tolerance (backup) the method of a uniform distribution of tasks between multiple network devices (e.g. servers) is used, which is called load balancing.

When a new task appears the software realizing balance should decide on which computing node the calculations associated with a new task should be performed. In addition, the balancing involves the transfer of parts of computing from the most loaded compute nodes to less loaded ones. When performing tasks, processors exchange communication messages among themselves. In the case of low-cost communication, some processors may be idle while others are overloaded. Also, high costs of communication are inexpedient. Consequently, the balancing strategy should be such that the computing nodes were loaded uniformly enough, but also the communication environment should not be overloaded.

In the scientific literature a lot of attention is paid to the load control in the distributed systems. Theoretical studies and development of fundamentals of the load distribution, creation of the mathematical apparatus, models and management methods for load balancing in the distributed systems were considered by such scientists as Hisao Kameda, Lie Li, Chonggun Kim, Yongbing Zhang [1], H. Mehta, P. Kanungo, M. Chandwani [2], V. Cardellini, M. Casalicchio, E. Casalicchio [3,4], Y.S. Hong, J.H. No, S.Y. Kim [5], Martin Randles, David Lamb, A. Taleb-Bendiab [6], Nayandeep Sran, Navdeep Kaur [7]. Such scientists as S. Keshav [8], O. Elzeki, M. Reshad, M. Elsoud [9] were developing and improving load balancing algorithms, depending on the task completion time on the computer. Brute force algorithms and those based on statistical data were considered by such scholars as H. Chen, F. Wang, N. Helian, G. Akanmu [10], Y. Hu, R. Blake, D. Emerson [11], Jing Liu, Xing-Guo Luo, Xing-Ming Zhang, Fan Zhang, Bai-Nan Li [12], Shamsollah Ghanbari, Mohamed Othman [13]. Algorithms based on biological phenomena were considered by Ratan Mishra, Anant Jaiswal [14], Dhinesh Babu L.D., P. Venkata Krishna [15].

The problem of the computational load balancing of the distributed application appears for the following reasons:

–the structure of the distributed application is non-uniform, various logical processes require different computing capacities;

– the structure of the computing complex (e.g. the cluster) is non-uniform, i.e. different computing nodes have different performance;

– the structure of inter-node communication is non-uniform because links connecting nodes may have different bandwidth characteristics.

The purpose of this work is to study the basic load balancing algorithms of distributed systems, their classification, on the basis of which it is possible to make representation analysis of algorithms with the description of their advantages and shortcomings, as well as a comparative analysis of the basic algorithms for different performance metrics.

## II. Classification of load balancing algorithms

The load balancing helps in fair allocation of computing resource to achieve a high user satisfaction and proper resource utilization. High resource utilization and a proper load balancing help in minimizing the resource consumption. It helps in implementing fail over, scalability, and avoiding bottlenecks [16, 17].

The load balancing is a technique that helps networks and resources by providing a maximum throughput with minimum response time. The load balancing divides the traffic between all servers, so data can be sent and received without any delay with the load balancing.

The major goals of the load balancing algorithms are:

- Cost effectiveness: The load balancing helps to provide better system performance at lower cost.

- Scalability and flexibility: The system for which the load balancing algorithms are implemented may be changed in size after some time. In this way the algorithm must handle situations of these types' and must be flexible and scalable.

- Priority: Prioritization of the resources or jobs needs to be done. So higher priority jobs get a better chance to be executed.

Load balancing algorithms can be classified into several types [17, 23]:

A. *Depending on the system state*

   *a) Static algorithm*

The current state of the node is not taken into account, a prior knowledge base is required about each node statistics and user requirements, not flexible, not scalable, not compatible with changing users' requirements and load. It is used in a homogeneous environment.

   *b) Dinamic algorithm*

This type of algorithms operates according to the dynamic changes in the nodes state, i.e., these algorithms collect, store and analyze information about the system state. Because of this, they are characterized by large loads balancing overhead. It is necessary to take into account the location of the processor to which the load is transferred by an overloaded CPU, the load estimation, limitation of the migrations number. If some node has failed, it will not stop the work of the entire network, but it will influence the system performance. It is used in a heterogeneous environment.

B. *Depending on who initiated the process*

- Initiated by a sender: a sender defines that the number of nodes is large and a sender initiates the execution of the load balancing algorithm.

- Initiated by a receiver: the load balancing requirement can be identified by a receiver/server in the distributed system, and then the server initiates the load balancing algorithm execution.

- Symmetric: This is a combination of the sender initiated and receiver initiated types.

The dynamic load balancing approach is divided into two types: distributed and undistributed (centralized) approaches. They are defined in the following way:

   *a) in the non-distributed approach:*

one node or group of nodes are responsible for the management and distribution of service information throughout the system. Other nodes do not distribute tasks and do not operate management functions, therefore this type of algorithms is not fault tolerant, the central decisions making processor can be overloaded. They are useful in small networks with a low load;

   *b) in the distributed approach:*

each node builds its load vector independently. All processors in the network are responsible for the load distribution and content of their own local database for taking effective decisions of load balancing. This causes large communication costs and complexity of algorithms. They are useful in large and heterogeneous networks.

The load balancing can have two forms at the distributed approach: cooperative and non-cooperative ones. At the cooperative approach the nodes operate side-by-side for a common goal, such as the overall response time perfection. At the non-cooperative approach the node operates independently of the purpose, such as perfection of the execution time of local problems.

The nodes are constantly interacting with each other and generate more messages at the distributed approach. Sending messages between the nodes to share information about updates of the system can cause the decrease in the system performance [1]. One node or a group of nodes solves the problem of the load balancing at the undistributed approach, which obtains two forms, namely: centralized and semidistributed ones.

In the centralized algorithms one unit is exclusively responsible for the load balancing of the whole system and it is called the central node. It is used in small networks. In the semi distributed algorithms a cluster is formed by the system group of nodes, the load balancing occurs in each cluster by the centralized type. The central node initiates the load balancing among the group of nodes in the cluster. Semidistributed algorithms exchange a lot of messages compared to the centralized ones.

The algorithm for selection of communication tasks defines the task to transfer. When selecting a task to send the algorithm can take into account that the overhead associated with sending the tasks should be minimal, the complexity of the task should be large, the number of links in the problem with local resources should be minimal.

The algorithm of choosing a partner is responsible for the selection of a suitable node for the balancing operation. A common technique in the distributed algorithms is nodes polling. The poll can be serial or parallel, it can use the results of the previous polls.

In the centralized algorithms the node applies to a specialized coordinator to determine a suitable partner for balancing. The coordinator collects and maintains up to date information about system components capacity, and the partner search algorithm uses this information.

The data collection mechanism collects information about the system capacity, defines sources of the information, the time of capacity data collection, storage space. There are several classes of the data collecting mechanism.

- Data collection on the need. In this class the distributed algorithms are used to collect information about the capacity when a node needs the load balancing.

- Periodic data collection. The algorithms of this class can be both centralized and distributed. Depending on the data collected, the algorithm initiates the load balancing.

- Data collection on change of state. In systems that implement algorithms of this class, the nodes themselves are spreading information about changing capacity when changing the internal state. In the case of the distributed algorithms data are sent to neighboring nodes, in the case of the centralized algorithms data are sent to the coordinator.

Classification of the load balancing algorithms is shown in Figure 1.

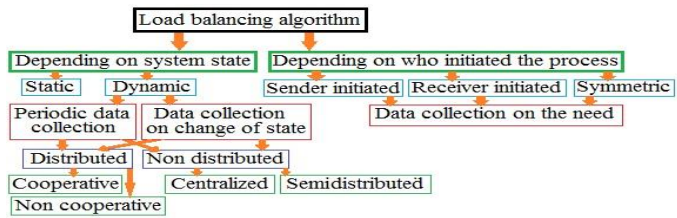

Fig. 1. Classification of load balancing algorithms

### III. ANALISIS OF LOAD BALANCING ALGORITHMS

There are many algorithms for load balancing, due to which higher throughput and improved response time in distributed systems is achieved. Each algorithm has both advantages and shortcomings [9-23].

Table 1 presents a comparative analysis of different load balancing algorithms by various efficiency metrics.

The efficiency of the load balancing algorithms is determined by several indicators, which are listed below [20-21]:

*a) Throughput*

This indicator is used to assess the total number of tasks that are successfully completed. High bandwidth is required for the overall system performance.

*b) Overhead*

Overhead is associated with the operation of any load-balancing algorithm, and it indicates the cost of the processes involved in the task and redistribution process. Overhead should be as low as possible.

TABLE I. COMPARATIVE ANALYSIS OF DIFFERENT LOAD BALANCING ALGORITHMS ON VARIOUS EFFICIENCY INDICATORS

| Algorithms/ Metrics | Throughput | Overhead | Fault tolerance | Migration time | Response time | Resource Utilization | Scalability | Performance |
|---|---|---|---|---|---|---|---|---|
| Task Scheduling based on LB [13, 18, 21, 22] | + | + | - | + | + | + | + | + |
| Opportunistic LB (OLB) [18, 21-22] | + | + | + | - | - | + | + | + |
| Round Robin [8, 18, 21] | + | + | - | - | + | + | - | + |
| Weighted Round Robin [8, 18, 23] | + | + | + | + | - | + | - | - |
| Randomized [18-19] | + | + | + | - | - | + | - | + |
| Min-Min [9, 10, 17-18, 21] | + | + | - | - | + | + | - | + |
| Max-Min [9, 16-18, 21] | + | + | - | - | + | + | - | + |
| Honeybee Foraging Behavior [18, 21] | + | - | - | - | - | + | + | + |
| Active Clustering [18, 23] | - | + | - | + | - | + | - | - |
| Compare and Balance [11, 18, 23] | + | + | + | + | + | - | + | + |
| Lock-free multiprocessing solution for LB [12, 18, 21] | - | + | + | - | - | - | - | - |
| Ant Colony Optimization [14-16, 18] | + | + | - | - | - | + | + | + |
| Shortest Response Time First [12, 17, 22] | - | + | - | + | + | + | - | + |
| Based Random Sampling [20, 22] | + | - | - | - | - | - | - | + |
| The two phase scheduling LB [13, 16, 22] | + | + | - | - | + | + | - | + |
| Active Clustering LB [11, 16, 22] | - | + | - | + | - | + | - | - |
| ACCLB [22] | + | + | - | - | - | + | + | + |
| Decentralized content aware [22] | + | + | - | + | + | - | + | + |
| Server-based LB [22] | - | - | + | + | + | + | + | + |
| Join-Idle-Queue [22-23] | + | + | - | + | + | - | + | + |
| Token Routing [19] | - | + | - | + | - | - | + | - |
| Central queuing [18] | - | + | - | + | + | + | - | + |
| Connection mechanism [12, 19, 23] | + | + | + | - | - | + | + | - |

*c) Fault tolerance*

This indicator measures the ability of the algorithm to perform the load balancing uniformly in the event of any failure. The good load balancing algorithm must be very insensitive to faults.

*d) Migration time*

It is defined as the total time of transition of a task from one node or resource to another. It should be minimized.

*e) Response time*

It is measured as the time interval between sending a request and receiving a response. It should be minimized in order to improve overall performance.

*f) Resource utilization*

The indicator is used to ensure the appropriate harnessing of all the system resources. This indicator must be optimized for efficiency of the load balancing algorithm.

*g) Scalability*

It is the ability of the algorithm to perform an uniform load balancing in the system according to the requirements upon increasing the number of nodes. The preferred algorithm is highly scalable.

*h) Performance*

It may be defined as the system efficiency. This indicator should be improved at a reasonable overhead, for example, reducing the response time and preserving an allowable delay.

## CONCLUSION

The most used load balancing algorithms of distributed systems are classified according to different types in this work. Based on the performed analysis of classification types of the load balancing algorithms the scope of each type of algorithms is indicated, the algorithms types necessary operation requirements are defined, defaults of each type of algorithms are shown. In this way one can select a particular type of the load balancing algorithm based on the specifics of a particular project or executable task, and the goals to be achieved. The description of the main features of load balancing algorithms, analysis of their advantages and defaults are also presented in this work. A comparative analysis of different load balancing algorithms on various performance metrics is carried out, i.e., the efficiency indicators are shown for each algorithm used in it. It is planned to realize a comparative analysis of the load balancing algorithms with different capacity in a variety of distributed systems: cloud technology, cluster systems, grid systems.